\documentclass[graybox]{svmult}

\usepackage{mathptmx}       
\usepackage{helvet}         
\usepackage{courier}        
\usepackage{type1cm}        
\usepackage{makeidx}         
\usepackage{graphicx}        
\usepackage{multicol}        
\usepackage[bottom]{footmisc}

\usepackage[latin2]{inputenc}
\usepackage{amsmath}
\usepackage{amssymb}
\usepackage{amsfonts}
\usepackage{stmaryrd}
\usepackage{theorem}
\usepackage{epstopdf}
\usepackage[parfill]{parskip} 
\usepackage{hyperref}
\hypersetup{pdftex,colorlinks=true,allcolors=blue}
\usepackage{hypcap}
\usepackage{algorithm}
\usepackage{algpseudocode} 
\usepackage{authblk}
\usepackage{tabto}
\usepackage{wrapfig} 
\usepackage{siunitx} 

\newtheorem{prb}{Problem}[section]

\makeindex             

\newcommand*{\vr}{\ensuremath{\varrho}}
\newcommand*{\ve}{\ensuremath{\varepsilon}}

\newcommand*{\vd}{\ensuremath{\delta}}

\newcommand*{\vs}{\ensuremath{\sigma}}

\newcommand*{\vo}{\ensuremath{\omega}}
\newcommand*{\vg}{\ensuremath{\gamma}}

\newcommand*{\R}{\ensuremath{\mathbb{R}}}

\newcommand*{\mto}{\ensuremath{\rightarrow}}

\newcommand{\libeq}{\mathrel{\mathop:}=} 

\newcommand*{\kon}{\ensuremath{k_{\mathrm{on}}}}
\newcommand*{\koff}{\ensuremath{k_{\mathrm{off}}}}
\newcommand*{\Kd}{\ensuremath{K_{\mathrm{d}}}}

\newcommand*{\dcdt}{\ensuremath{\partial_t c}}

\newcommand*{\dcdx}{\ensuremath{\partial_x c}}
\newcommand*{\ddcdx}{\ensuremath{\partial_x^2 c}}

\newcommand*{\Leq}{\ensuremath{\bar{L}}}
\newcommand*{\Teq}{\ensuremath{\bar{T}}}

\newcommand*{\Ceq}{\ensuremath{\bar{C}}}

\newcommand*{\Tini}{\ensuremath{T_{\mathrm{ini}}}}

\newcommand*{\tmax}{\ensuremath{t_{\mathrm{max}}}}
\newcommand*{\xdet}{\ensuremath{x_{\mathrm{det}}}}

\newcommand*{\mS}{\mathrm{S}}

\newcommand*{\mIC}{\mathrm{IC}}

\newcommand*{\mmolmc}{\ \si{\mole/\cubic\metre}}

\begin{document}

\title*{A Computational Resolution of the Inverse Problem of Kinetic Capillary Electrophoresis (KCE)}
\titlerunning{A Computational Resolution of the Inverse Problem of KCE}
\author{J\'{o}zsef Vass and Sergey N. Krylov}
\authorrunning{J. Vass and S.N. Krylov}
\institute{J\'{o}zsef Vass \at Centre for Research on Biomolecular Interactions, Department of Chemistry, York University, Toronto, ON, M3J 1P3, Canada, \email{jvass@yorku.ca}
\and Sergey N. Krylov \at Centre for Research on Biomolecular Interactions, Department of Chemistry, York University, Toronto, ON, M3J 1P3, Canada, \email{skrylov@yorku.ca}}
%
%
\maketitle

\abstract*{Determining kinetic rate constants is a highly relevant problem in biochemistry, so various methods have been designed to extract them from experimental data. Such methods have two main components: the experimental apparatus and the subsequent analysis, the latter often dependent on mathematical theory. Thus the theoretical approach taken influences the effectiveness of constant determination. A computational inverse problem approach is hereby presented, which does not merely give a single rough approximation of the sought constants, but is inherently capable of determining them from exact signals to arbitrary accuracy. This approach is thus not merely novel, but opens a whole new category of solution approaches in the field, enabled primarily by an efficient direct solver.}

\abstract{Determining kinetic rate constants is a highly relevant problem in biochemistry, so various methods have been designed to extract them from experimental data. Such methods have two main components: the experimental apparatus and the subsequent analysis, the latter often dependent on mathematical theory. Thus the theoretical approach taken influences the effectiveness of constant determination. A computational inverse problem approach is hereby presented, which does not merely give a single rough approximation of the sought constants, but is inherently capable of determining them from exact signals to arbitrary accuracy. This approach is thus not merely novel, but opens a whole new category of solution approaches in the field, enabled primarily by an efficient direct solver.}
\ \\
\section{Introduction} \label{s01}

\subsection{Aims and Overview} \label{s0101}

The direct problem of efficiently generating accurate numerical solutions to the set of partial differential equations of Kinetic Capillary Electrophoresis (KCE) \cite{krylov2007kinetic}, has been resolved earlier via a multimesh algorithm \cite{vasskrylov2016ccdr}, which fully overcomes the typical instability arising from the interaction of the diffusion and convection terms. This potent solution to the direct problem allows the effective resolution of the inverse problem on a reasonable timescale, as we shall hereby discuss and demonstrate. The inversion essentially entails the optimization of a non-linear error objective function, computed between the experimental target signal and the signals generated numerically at each iteration. Estimating the starting point for the optimization, posed a challenge in itself \cite{vasskrylov2017kceparest}. Furthermore, control of the error in the sought parameters is demonstrated, enabling the resolution of this inverse problem to arbitrary accuracy for exact target signals. This work has also been implemented as a software package \cite{so002}.

The KCE system of equations includes kinetic rate constants in the reaction term, which can be viewed as parameters that the solution functions of this system of partial differential equations depend on. The aim of the inverse problem is to find or approximate certain parameters -- the rate constants of complex formation and dissociation -- that induce an a priori given exact solution. If such a method can arbitrarily approximate an exact solution, then it can be applied to experimental data reliably. Therefore, the goal becomes to minimize an error function between the given solution and the approximating solutions, which must be generated for each set of test parameters as the optimization progresses. The generation of such parametric solutions is essentially akin to a function evaluation in an optimization iteration step, which must clearly be both efficient and accurate.

Various experimental approaches have been taken in the literature to approximating rate constants, though not to arbitrary accuracy, which differentiates our method in its novelty. Existing methods include \cite{wilson2002analyzing, hornblower2007single, al2005fluorescence, li2007kinetics, abdiche2008determining, rich2007higher}, while KCE-based methods are surveyed in the previous article \cite{vasskrylov2017kceparest}.

\subsection{The Physical Model} \label{s0102}

We adopt the physical model and the notations covered in earlier articles \cite{vasskrylov2016ccdr, vasskrylov2017kceparest}, originally introduced in \cite{krylov2007kinetic, berezovski2002nonequilibrium, petrov2005kinetic}.

The concentration vector of reactants $c = (L,\ T,\ C):\R_+^2\mto\R_+^3$ is a mapping defined over spacetime points $(t,\ x)\in [0,\ \tmax]\times [0,\ \xdet]$. It satisfies the equation
\[ \dcdt + v\cdot\dcdx = D\cdot\ddcdx + R(c) \]
where $v = (v_L,\ v_T,\ v_C)\in\R_+^3$ and $D = (D_L,\ D_T,\ D_C)\in\R_+^3$, and $\cdot$ denotes the Hadamard product. The reaction term takes the form
\[ R(c) = (-\kon LT + \koff C,\ \ -\kon LT + \koff C,\ \ \kon LT -\koff C): \R_+^2\mto\R^3 \]
where $k = (\kon,\ \koff)\in\R_+^2$. Lastly, define $\Kd\libeq\koff/\kon$.

Since as it will become apparent in Section \ref{s03}, the scope of this paper is constrained by the parameter estimation methods introduced earlier \cite{vasskrylov2017kceparest}, we only consider the NECEEM initial and boundary conditions for the above partial differential equations. The initial conditions are given by $\mIC(x) = c(0,\ x) = \bar{c}\cdot\vr(x/l)$, where $\bar{c}= (\Leq,\ \Teq,\ \Ceq)\in\R_+^3$ (note: $\Kd = \Leq\Teq/\Ceq$) and $\vr:\R_+\mto\R_+^3$ is a vector of asymmetric Gaussian density functions, and $l > 0$. The left boundary condition is $c(t,\ 0) = 0$, while the right one is $\dcdx(t,\ \xdet) = 0$, for computational purposes.

The signal is defined as the function
\[ \mS[k,\ \vg](t) \libeq (L+C)(t,\ \xdet),\ \ t\in [0,\ \tmax] \]
parametrized by the above $k$ and the asymmetric Gaussian plug parameters
\[ \vg = (\mu_L,\ \vs_L^1,\ \vs_L^2,\ h_L,\ \mu_T,\ \vs_T^1,\ \vs_T^2,\ h_T,\ \mu_C,\ \vs_C^1,\ \vs_C^2,\ h_C) \]
which denote the center, the left and right standard deviations, and the height of the Gaussian initial conditions (dependent on $\bar{c}$).

See our previous articles for further details \cite{vasskrylov2017kceparest, vasskrylov2016ccdr, krylov2007kinetic}, such as the physical meaning of the above constants.

\section{The Inverse Problem of KCE} \label{s03}

\subsection{Problem Statement} \label{s0301}

The direct problem of generating a solution to the KCE equations, introduced above, can be inverted to inquire what parameters $(k_*,\ \vg_*)\in\R_+^{14}$ induced a given signal $S_*:[0,\ \tmax]\mto\R_+$, meaning $S_*=S[k_*,\ \vg_*]$ by the earlier notations, which we refer to as the ``target signal''. Since as we shall see in the next section, the values of $k_*$ and $\vg_*$ are not independent, we will only need to invert over some of these parameters, denoted by $\vo\in\R_+^{10}$, while updating the definition of the signal mapping $S[\vo]$ to eliminate the redundant parameters.

Define the error function as $E(\vo)\libeq D(S_*,\ S[\vo])\ \ (\vo\in\R_+^{10})$ where $D$ is any metric, typically the Euclidean. This $E$ is the target function to be minimized during inversion, below some required threshold $\ve >0$. The problem is thus the following.

\begin{prb} \label{s030101} \textup{(KCE Inverse Problem)}\ \
Given a threshold $\ve > 0$ and an $S_*:[0,\ \tmax]\mto\R_+$ function, induced by unknown KCE parameters $\vo_*$ and defined as $S_*\libeq S[\vo_*]$, find an $\vo$ parameter vector such that $E(\vo)< \ve$.
\end{prb}

This inverse problem may appear at first to be ill-posed, meaning its solution is not necessarily unique, and $E(\vo)= 0$ doesn't trivially imply that $\vo = \vo_*$. The signal mapping $S$ is not only a superposition, but also a slice of a surface, resulting in a significant loss of information relative to the full solution vector $c = (L,\ T,\ C)$. However, our computational experiments strongly suggest well-posedness, making it a worthwhile conjecture, equivalent to asserting that the error function possesses a unique minimum.

\subsection{The Optimization Space} \label{s0302}

In this section, we clarify how the optimization space over $\vo\in\R_+^{10}$ reduces from the seemingly straightforward variables $(k,\ \vg)\in\R_+^{14}$. As mentioned, this reduction is partly necessitated by the interdependencies between the coordinates of the latter vector, which originate in the physicochemical model of KCE \cite{krylov2007kinetic}.

Another reason for the reduction, is that the variation in the $T$-plug parameters in $\vg$ does not really affect the simulated signal. The standard deviations are not particularly relevant for the effectiveness of the computational inversion. The quantity of the $T$ substance in this chemical reaction can be controlled solely via the height of the $T$-plug, while the plug center may be allowed to vary within $1-2$ orders of magnitude. Though the height is in fact dependent on some other parameters, as stated below.

Furthermore, it must be noted that $\kon$ typically has little bearing on the NECEEM signal, as reasoned in earlier papers \cite{vasskrylov2017kceparest, okhonin2004nonequilibrium}, so the optimization process may place a greater emphasis on optimizing in $\koff$.

The full vector of parameters
\[ (k,\ \vg) = (\kon,\ \koff,\ \mu_L,\ \vs_L^1,\ \vs_L^2,\ h_L,\ \mu_T,\ \vs_T^1,\ \vs_T^2,\ h_T,\ \mu_C,\ \vs_C^1,\ \vs_C^2,\ h_C) \]
is reduced to the vector of optimization parameters
\[ \vo = (\koff,\ \mu_L,\ \vs_L^1,\ \vs_L^2,\ h_L,\ \mu_T,\ \mu_C,\ \vs_C^1,\ \vs_C^2,\ h_C). \]

To make up for the missing coordinates $\kon,\ \vs_T^1,\ \vs_T^2,\ h_T$, which are still necessary to generate signals $S[\vo]$ for the calculation of the error $E$, some relationships between the coordinates must be observed. Firstly
\[ \Leq = \sqrt{2\pi}\ \frac{h_L}{l}\ \frac{\vs_L^1 + \vs_L^2}{2} \]
and similarly for $\Ceq$, while $\Teq = 1000\Tini - \Ceq$ where $\Tini$ is the initial pre-equilibrium concentration of the $T$-plug. Its height can then be calculated as
\[ h_T = \frac{l\Teq}{\sqrt{2\pi}}\ \big/\ \frac{\hat{\vs}_T^1 + \hat{\vs}_T^2}{2} \]
where $\hat{\vs}_T^{1,2}$ are the standard deviation estimates for the asymmetric Gaussian $T$-plug \cite{vasskrylov2017kceparest}. Using mere estimates does not significantly affect the effectiveness of inversion, as reasoned heuristically above, i.e. the minimization of the error function $E$ below a given threshold. Lastly, $\ \kon = \koff\Ceq/(\Leq\Teq)$ which is simply the relationship mentioned in Section \ref{s0102}.

Lastly, we remark that the inversion using any optimization algorithm we tested, proved to be significantly more efficient when carried out in logarithmic space, in all independent variables.

\subsection{Error Control} \label{s0303}

As stated in the KCE Inverse Problem \ref{s030101}, the inversion is formulated as the minimization of the error function $\vo\mto E(\vo)$ below some threshold $\ve > 0$, where the signal error is calculated as $E(\vo)= D(S_*,\ S[\vo])$, typically with the Euclidean metric $D$ and some target signal $S_* = S[\vo_*]$.

It remains unclear, how this minimization can be made practical. For the purpose of scientific applications, our inevitable aim must be to ensure that the error in the $k = (\kon,\ \koff)$ parameter is minimized below a threshold $\vd > 0$, prescribed a priori. Therefore, the task becomes to determine -- or estimate -- what $\ve$ threshold above is necessary to ensure this $\vd$.

The matter can be resolved under the rather weak hypothesis that there is a local Lipschitz constant $L$ satisfying
\[ d(k_*,\ k)\leq d(\vo_*,\ \vo)\leq L\cdot D(S[\vo_*],\ S[\vo]) = L\cdot E(\vo) \]
in a neighborhood of the point $\vo_*$ in the optimization space. Since this point is unknown, computationally we ensure a large enough neighborhood in logarithmic space around our estimate -- derived earlier \cite{vasskrylov2017kceparest} -- likely to contain the sought $\vo_*$. Typically, the neighborhood is taken within two orders of magnitude, which proved to be sufficient according to our computational experiments.

The Lipschitz constant $L$ can then be estimated within this neighborhood, by taking random pairs of points $\vo_1,\ \vo_2$ in it, and taking the maximum of the ratios $d(\vo_1,\ \vo_2)/D(S[\vo_1],\ S[\vo_2])$.

Using this estimated $L$, the threshold in the signal error $E(\vo)$ must be taken to be $\vd\libeq\ve/L$, in order to ensure that the error in $k$ falls below the prescribed signal threshold $\ve >0$.

\subsection{Implementation} \label{s0304}

The inverse solver \cite{so002} is essentially a non-linear optimization process, which minimizes the error function introduced in Section \ref{s0301}. At each evaluation of this function, the direct solver must be called to generate a signal $S[\vo]$ for the current iteration of parameters $\vo$. Thus the runtime of the inverse solver is fundamentally implied by that of the direct solver. The stability and efficiency of the direct solver was established in our earlier article \cite{vasskrylov2016ccdr}.

Finding a well-functioning -- or perhaps even ``ideal'' -- optimization algorithm, was in itself a challenge, and we have tested several. The \textbf{fmincon} MATLAB function with the \textit{interior-point} algorithm proved to be the most robust and efficient. The \textit{bfgs} Hessian approximation option (a dense quasi-Newton approximation \cite{site006}) is typically sufficient, but occasionally the \textit{lbfgs} option (a limited-memory, large-scale quasi-Newton approximation \cite{site007}) must also be run, in case the \textit{bfgs} option fails to converge below the required signal error threshold within the allocated time. Other tested algorithms include Cuckoo Search \cite{yang2009cuckoo}, Flower Pollination \cite{yang2012flower}, and Harmony Search \cite{geem2001new}, each of which proved to be less robust than \textbf{fmincon}, but are nevertheless available in our package \cite{so002}.

Each inverter -- i.e. error minimization subroutine, with a particular optimization algorithm -- was tested for simulated signals induced by known parameters, and then compared to the output parameters in terms of relative error (see Figures \ref{s040101} and \ref{s040201}). In practice, the inverters are executed on experimental signals (from unknown parameters), in which case, the accuracy of the result can be gauged via the error control method described in Section \ref{s0303}, and demonstrated in Section \ref{s04}.

\section{Performance Analysis} \label{s04}

\begin{figure}[H]
\centering
\includegraphics[width=330pt]{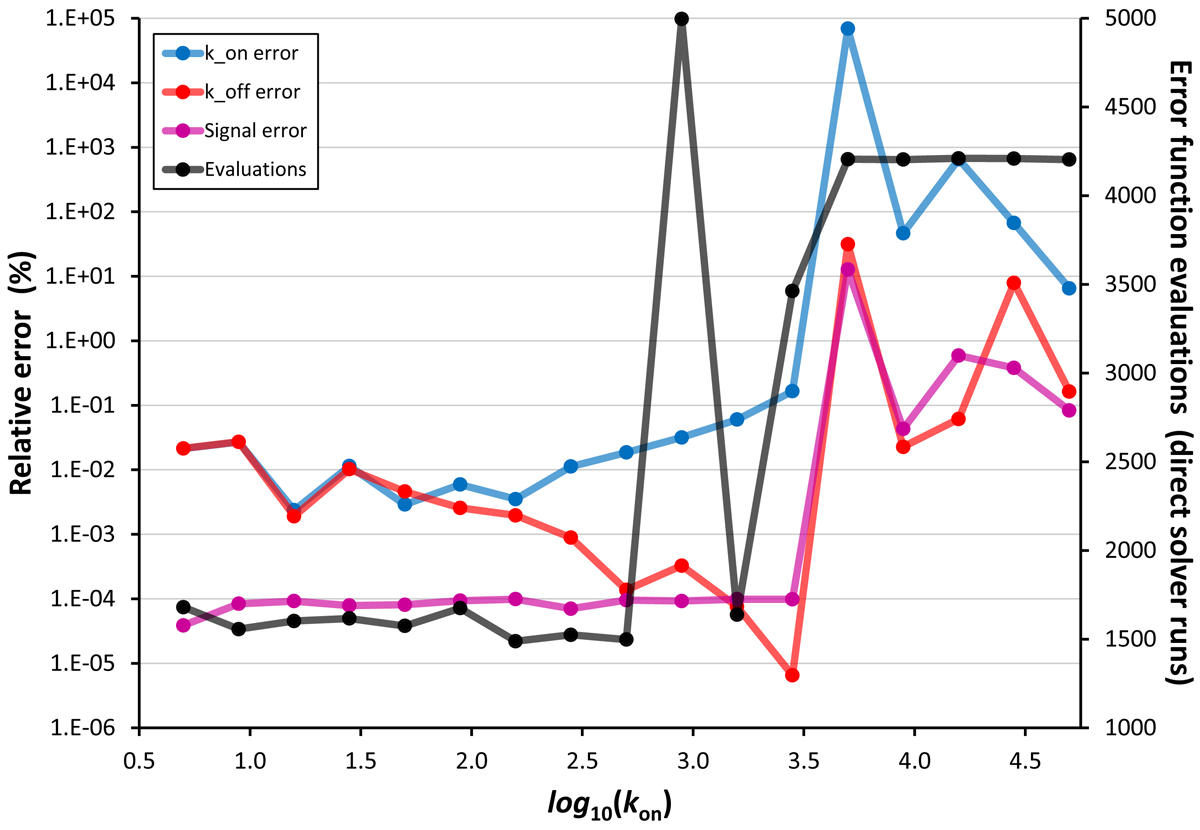}
\caption{Variation of the relative error in $k$ with respect to increasing $\log_{10}(\kon)$ values, at constant $\Kd = 2\times 10^{-6}\mmolmc$. The $t$-mesh size is $300$, and the error threshold is $\ve= 0.0001\%$.} \label{s040101}
\end{figure}

Figure \ref{s040101} depicts the performance analysis of the primary inverter, running the BFGS and L-BFGS \cite{site006, site007} interior point optimization algorithms of MATLAB (\textbf{fmincon} with the \textit{interior-point} algorithm, and Hessian approximation methods \textit{bfgs} and \textit{lbfgs}). The target threshold of $0.0001\%$ in the signal error tends to ensure two correct decimal places in $k$ in scientific notation. This threshold is reached by the inverter up to $\log_{10}(\kon)\approx 3.45$ on the horizontal axis, but it fails to converge above that value. The ca. $4200$ evaluations is the default upper iteration bound that each optimization algorithm is allowed to run, at this $t$-mesh size. The ca. $5000$ evaluations at ca. $2.75$ results from running the inverter twice, trying both settings (\textit{bfgs} and \textit{lbfgs}).

\begin{figure}[H]
\centering
\includegraphics[width=330pt]{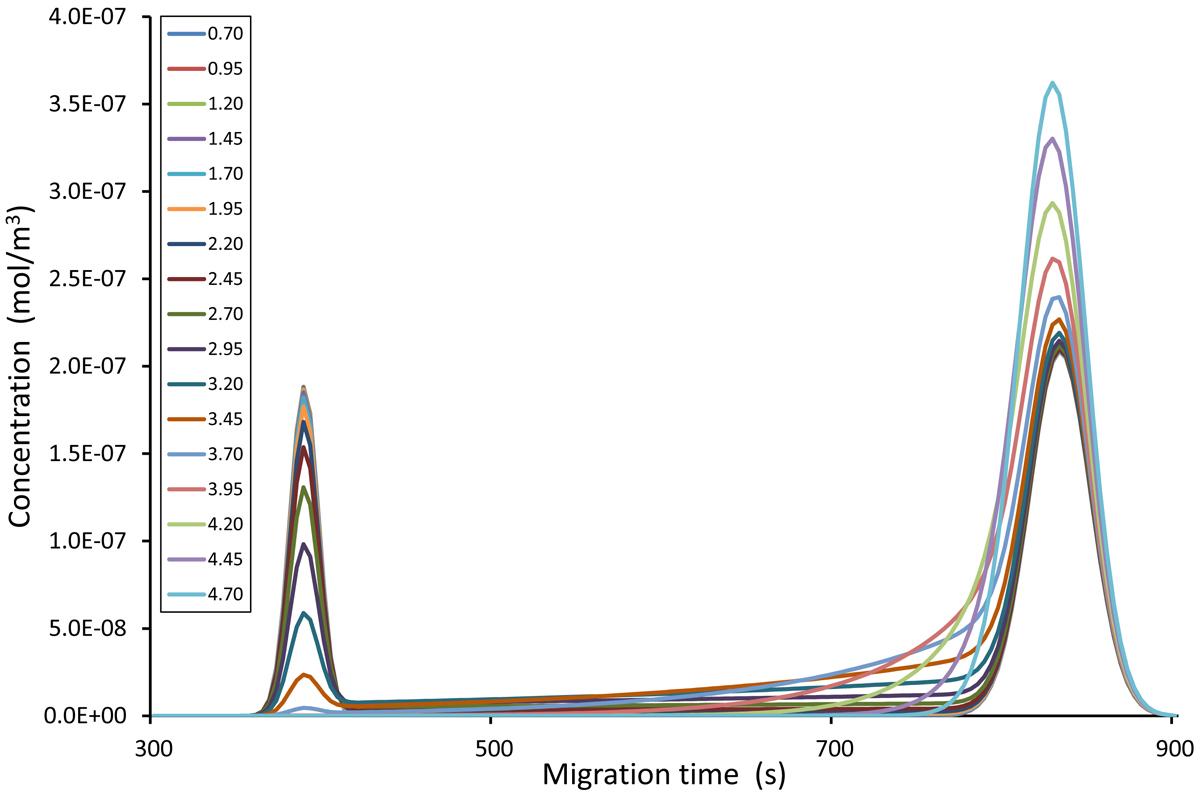}
\caption{Simulated electropherogram signals for increasing $\log_{10}(\kon)$, at a $t$-mesh size of $300$.} \label{s040102}
\end{figure}

Figure \ref{s040102} accompanies Figure \ref{s040101}, for reference. Apparently for $\log_{10}(\kon)> 3.45$ the $C$-peak vanishes, visually elucidating the divergence of the inverter. The necessity of a prominent $C$-peak thus becomes a practical rule of thumb for the reliable inversion of experimental signals.

\begin{figure}[H]
\centering
\vspace{-0.65cm}
\includegraphics[width=330pt]{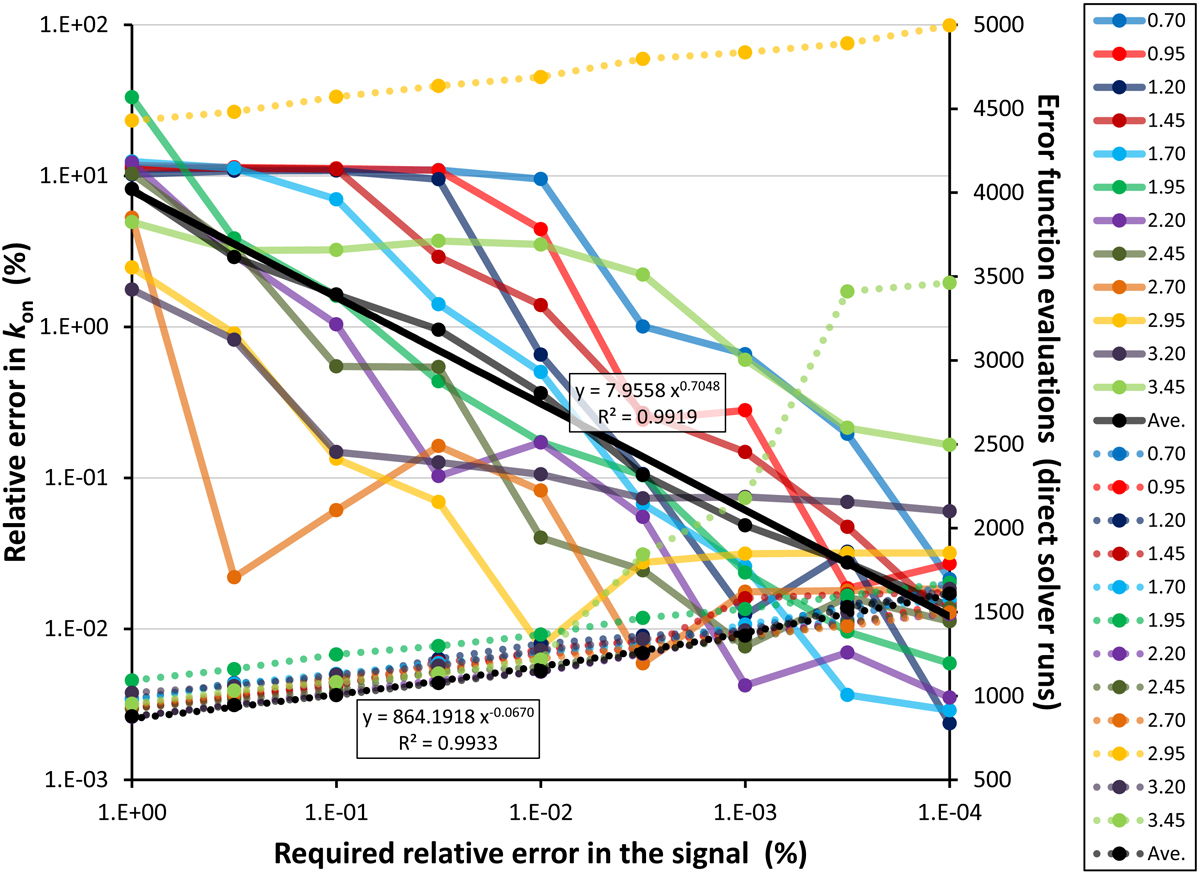}
\caption{Illustration of the correlation between the decrease in the optimization threshold in the signal relative error and the decrease in the $\kon$ relative error, for various $\log_{10}(\kon)$ cases (labeled in the legend). The corresponding runtimes are also plotted. Only the $\log_{10}(\kon)\le 3.45$ cases are plotted, where convergence of the inverter was ensured (see Figure \ref{s040101}).} \label{s040201}
\end{figure}

Figure \ref{s040201} is a log-log graph (with decreasing horizontal axis), which demonstrates a definite power law relationship on average, with an exponent of ca. $0.7$ between the optimization threshold in the signal relative error and the relative error between the original $\kon$ value and the one determined by the inverter. The error in $\koff$ tends to be close to the same value, or less, so it is not plotted. For lower $\kon$ values, the relative error in $k$ apparently stagnates for higher threshold errors in the signal, but does begin to decrease later. The demonstrated power law may be different for other experimental parameter sets.

Interestingly, the \textbf{fmincon} optimization subroutine also exhibits a power law relationship between its runtime and the error threshold in the signal, with a negligible exponent of ca. $-0.07$, not considering the outlier. In the outlier $\log_{10}(\kon)\approx 2.95$ case, the parameters appear to conspire so that only the \textit{lbfgs} option is able to tackle the inversion (the plotted runtimes include the failed attempts with the \textit{bfgs} option). Nevertheless, the overall linear relationship ensures that the runtime remains predictable for various thresholds.

\begin{figure}[H]
\centering
\includegraphics[width=330pt]{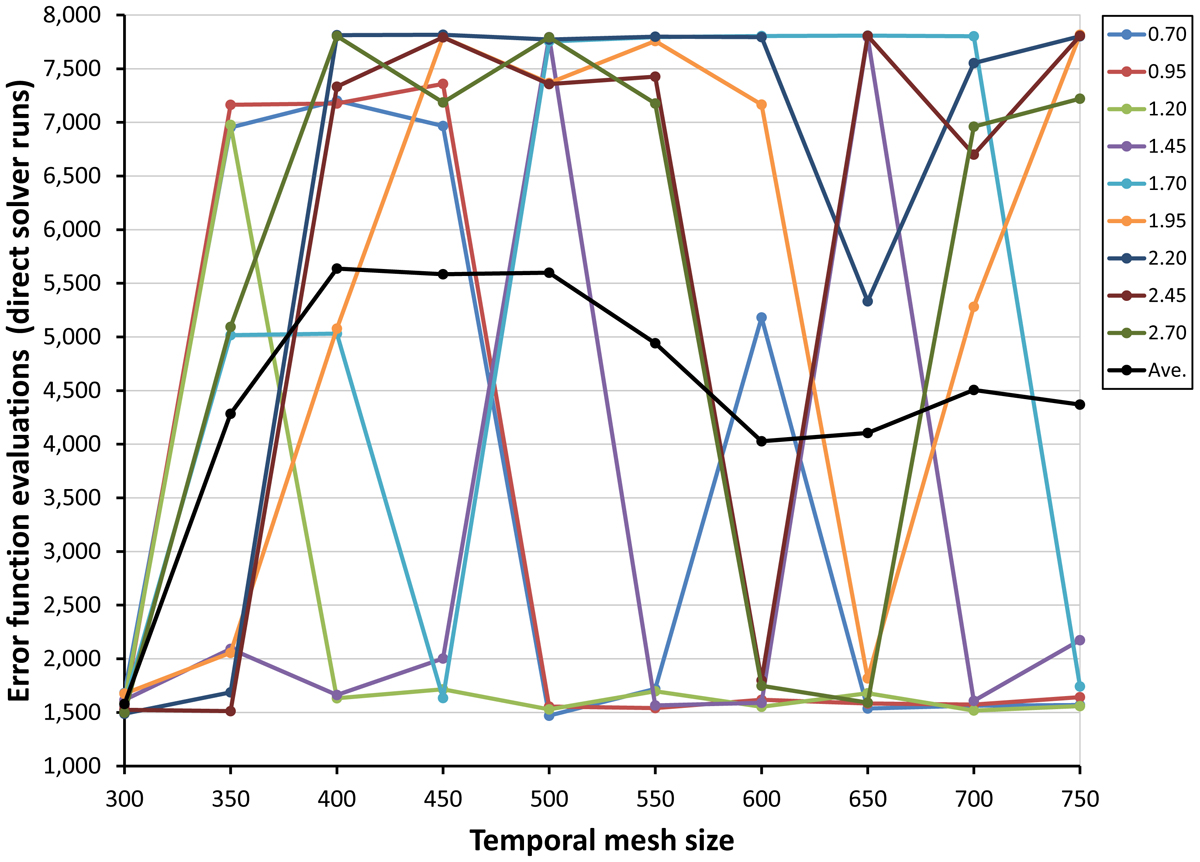}
\caption{Illustration of the variation of runtime with increasing temporal mesh size (to which the spatial is proportional). Only the $\log_{10}(\kon)\le 2.7$ cases are plotted, where convergence of the inverter was ensured within a reasonable timeframe for a temporal mesh size of $300$ (see Figure \ref{s040101}).} \label{s040301}
\end{figure}

Based on Figure \ref{s040301}, we can conclude that for all the $\log_{10}(\kon)\le 2.7$ test cases, and for all tested mesh sizes, the interior point inverter was successful. The case runtimes, however, did not follow a clear relationship with the increasing mesh size. Taking the average among all cases (in black), however, does exhibit a somewhat clear trend.

\newpage
\newpage

\section{Concluding Remarks} \label{s05}

This article aimed to resolve the inverse problem of Kinetic Capillary Electrophoresis, involving a set of partial differential equations, parameterized in the initial and boundary conditions, as well as the equations themselves. The problem was reformulated as the non-linear minimization of a certain error function, each evaluation of which required the generation of a solution with the direct solver.

The main challenge thus became to identify an optimization algorithm capable of carrying out this minimization to arbitrary accuracy, robustly and efficiently, which was accomplished. Furthermore, a local Lipschitz condition was utilized to relate the error in the signal to the error in the sought parameters, in order to control the accuracy in the latter. This is a definite novelty relative to earlier work on this topic.

While this article focused on the design of a computational method and its practical robust implementation, the implied theoretical questions nevertheless project avenues for future research. The most relevant problems being: (1) uniqueness of the global minimizer of the error function; (2) continuous differentiability of the error function. There is strong consistent computational evidence of a unique minimizer, according to our experiments. The second property would imply local Lipschitz continuity (hypothesized in Section \ref{s0303}), and validate the use of gradient-based optimization algorithms.

This work was supported by the Natural Sciences and Engineering Research Council of Canada (grant: CRDPJ 485321-15).

\bibliographystyle{abbrv}
\bibliography{mybib2}

\end{document}